\documentclass[pra,aps,showpacs,showkeys,reprint]{revtex4}
\usepackage{graphicx}
\usepackage{amsmath}
\usepackage{amssymb}

\begin{document}
\title{\bf The minimal length and the Shannon entropic uncertainty relation}
\author{Pouria Pedram}
\email[Electronic address: ]{p.pedram@srbiau.ac.ir}
\affiliation{Department of Physics, Science and Research Branch,
Islamic Azad University, Tehran, Iran}
\date{\today}

\begin{abstract}
In the framework of the generalized uncertainty principle, the
position and momentum operators obey the modified commutation
relation $[X,P]=i\hbar\left(1+\beta P^2\right)$ where $\beta$ is the
deformation parameter. Since the validity of the uncertainty
relation for the Shannon entropies proposed by Beckner,
Bialynicki-Birula, and Mycielski (BBM) depends on both the algebra
and the used representation, we show that using the formally
self-adjoint representation, i.e., $X=x$ and
$P=\tan\left(\sqrt{\beta}p\right)/\sqrt{\beta}$ where
$[x,p]=i\hbar$, the BBM inequality is still valid in the form
$S_x+S_p\geq1+\ln\pi$ as well as in ordinary quantum mechanics. We
explicitly indicate this result for the harmonic oscillator in the
presence of the minimal length.
\end{abstract}


 \maketitle

\section{Introduction}\label{sec1}
The existence of a minimal observable length proportional to the
Planck length $\ell_{\mathrm{P}}=\sqrt{G\hbar/c^3}\approx 10^{-35}m$
is motivated by various proposals of quantum gravity such as string
theory, loop quantum gravity, noncommutative geometry, and
black-hole \cite{felder,p10,p11}. Indeed, several schemes have been
established to investigate the effects of the minimal length range
from astronomical observations \cite{laser3,laser3-2} to table-top
experiments \cite{laser4}. In particular, a measurement method is
proposed recently to detect this fundamental length scale which is
based on the possible deviations from ordinary quantum commutation
relation at the Planck scale within the current technology
\cite{laser4}.

Deformed commutation relations have attracted much attention in
recent years and several problems range from classical to quantum
mechanical systems have been studied exactly or approximately in the
context of the Generalized (Gravitational) Uncertainty Principle
(GUP). Among these investigations in quantum domain, we can mention
the harmonic oscillator \cite{4,9,10}, Coulomb potential
\cite{13,pH1,pH2}, singular inverse square potential \cite{14}, coherent states \cite{coh},
Dirac oscillator \cite{11}, Lamb's shift, Landau levels, tunneling
current in scanning tunneling microscope \cite{25}, ultra-cold
neutrons in gravitational field \cite{24,24b}, Casimir effect
\cite{27}, relativistic quantum mechanics \cite{p1,p2,p3}, and
cosmological problems \cite{p4,Babak2,Babak3,Jalalzadeh}. On the
other hand, in the classical domain, deformed classical systems in
phase space \cite{301,302}, Keplerian orbits \cite{303}, composite systems \cite{304}, and the
thermostatistics \cite{305,306} have been investigated in the presence of the
minimal length.

In the last decade, many applications of information theoretic
measures such as entropic uncertainty relations, as alternatives to
Heisenberg uncertainty relation, appeared in various quantum
mechanical systems
\cite{q000,q00,q0,q1,q2,q3,q4,q5,q6,q7,q8,q9,q10,q11,q121,q122}. The
concept of statistical complexity was introduced by Claude Shannon
in 1948 \cite{Shannon} where the term uncertainty can be considered
as a measure of the missing information. In particular, it is shown
that the information entropies such as Shannon entropy may be used
to replace the well-known quantum mechanical uncertainty relation.
The first entropic uncertainty relation for position and momentum
observables was proposed by Hirschmann \cite{Hir} and is later
improved by Beckner, Bialynicki-Birula, and Mycielski (BBM)
\cite{Babenko,Beckner,bia1,bia2}.

In the context of the generalized uncertainty principle, there
exists two questions: (i) Are the wave functions in position space
and momentum space related by the Fourier transform in arbitrary GUP
algebra in the form $[X,P]=i\hbar f(X,P)$? (ii) If in a particular
algebra these wave function are not related by the Fourier
transform, is the BBM inequality still valid? Note that, the
validity of the BBM entropic uncertainty relation depends on both
the algebra and the used representation. For instance, consider the
modified commutation relation in the form $[X,P]=i\hbar(1+\beta
P^2+\alpha X^2)$ which implies a minimal length and a minimal
momentum proportional to $\hbar \sqrt{\beta}$ and
$\hbar\sqrt{\alpha}$, respectively. This form of GUP has no formally
self-adjoint representation in the form $X=x$ and $P=f(p)$. So the
momentum space and coordinate space wave functions are not related
by the Fourier transform in any representation and the BBM
uncertainty relation does not hold in this framework.

The well-known Robertson uncertainty principle for two non-commuting
observables is given by
\begin{eqnarray}\label{rob}
\Delta A\,\Delta B \geq \frac{1}{2}|\langle\psi|[A,B]|\psi\rangle|,
\end{eqnarray}
where $[A,B]$ denotes the commutator of $A$ and $B$ and $\Delta A$
and $\Delta B$ are their dispersions. However, this uncertainty
relation suffers from two serious shortcomings \cite{23,26}: (i) For
two non-commuting observables of a finite $N$-dimensional Hilbert
space, since the right-hand side of Eq.~(\ref{rob}) depends on the
wave function $\psi$, it is not a fixed lower bound. Indeed, if
$\psi$ is the eigenstate of the observable $A$ or $B$, the
right-hand side of Eq.~(\ref{rob}) vanishes and there is no
restriction on $\Delta A$ or $\Delta B$ by this uncertainty
relation. (ii) The dispersions cannot be considered as suitable
measures for the uncertainty of two complementary observables with
continuous probability densities. This problem is more notable when
their corresponding probability densities contain several sharp
peaks. Among various proposals for the uncertainty relations that
are not suffered from these shortcomings, we can mention the
information-theoretical entropy instead of the dispersions which is
a proper measure of the uncertainty.

In this paper, we study the effects of the minimal length on the
entropic uncertainty relation. In this scenario, the position and
momentum operators obey the modified commutation relation
$[X,P]=i\hbar\left(1+\beta P^2\right)$ where $\beta$ is the
deformation parameter. Using formally self-adjoint representation of
the algebra, we show that the coordinate space and momentum space
wave functions are related by the Fourier transform and consequently
the BBM inequality is preserved. However, as we shall show, in the
quasi-position representation the momentum space and quasi-position
space wave functions are not related by the Fourier transformation
and the BBM inequality does not hold. As an application, we obtain
the generalized Schr\"odinger equation for the harmonic oscillator
and exactly solve the corresponding differential equation in
momentum space. Then, we find information entropies for the two
lowest energy eigenstates and explicitly show the validity of the
BMM inequality in the presence of the minimal length.

\section{The Generalized Uncertainty Principle}
In one-dimension, the deformed commutation relation reads \cite{4}
\begin{eqnarray}\label{gupc}
[X,P]=i\hbar(1+\beta P^2),
\end{eqnarray}
which results in $\Delta X \Delta P \geq \frac{\hbar}{2}\left[1+
\beta(\Delta P)^2 \right]$ (generalized uncertainty principle) and
for $\beta\rightarrow0$ the well-known commutation relation in
ordinary quantum mechanics is recovered. Notice that, $\Delta X$
cannot take arbitrarily small values and the absolutely smallest
uncertainty in position is given by $(\Delta
X)_{min}=\hbar\sqrt{\beta}$.

Now, consider the \emph{formally} self-adjoint representation
\cite{10}
\begin{eqnarray}\label{x0p01}
X &=& x,\\ P &=&
\frac{\tan\left(\sqrt{\beta}p\right)}{\sqrt{\beta}},\label{x0p02}
\end{eqnarray}
where $[x,p]=i\hbar$,
$\frac{-\pi}{2\sqrt{\beta}}<p<\frac{\pi}{2\sqrt{\beta}}$, and it
exactly satisfies Eq.~(\ref{gupc}). In this representation, the
ordinary nature of the position operator is preserved and the inner
product of states takes the following form
\begin{eqnarray}\label{inner}
\langle\psi|\phi\rangle=\int_{-\frac{\pi}{2\sqrt{\beta}}}^{+\frac{\pi}{2\sqrt{\beta}}}\mathrm{d}p\,\psi^{*}(p)\phi(p).
\end{eqnarray}

Note that, although the position operator obeys
$X^{\dagger}=X=i\hbar\partial/\partial p$ in momentum space, we have
${\cal D}(X)\subset {\cal D}(X^{\dagger})$. So, $X$ is merely
symmetric and it is not a true self-adjoint operator. However, based
on the von Neumann's theorem, for the momentum operator we obtain
$P=P^\dagger$ and ${\cal D}(P)={\cal
D}(P^{\dagger})=\left\{\phi\in{\cal
D}_{max}\left(\mathbb{R}\right)\right\}$ \cite{10}. Thus, $P$ is
indeed a self-adjoint operator. Moreover, the scalar product and the
completeness relation read
\begin{eqnarray}\label{comp1}
\langle p'|p\rangle= \delta(p-p'),\\
\int_{-\frac{\pi}{2\sqrt{\beta}}}^{+\frac{\pi}{2\sqrt{\beta}}}\mathrm{d}
p\, |p\rangle\langle p|=1.\label{comp2}
\end{eqnarray}

In momentum space, the eigenfunctions of the position operator are
given by the solutions of the eigenvalue equation
\begin{eqnarray}
X\,u_x(p)=x\,u_x(p).
\end{eqnarray}
Here, $u_x(p)=\langle p|x\rangle$ which can be expressed as
\begin{eqnarray}\label{ux}
u_x(p)=\frac{1}{\sqrt{2\pi}}\exp\left({-\frac{i p}{\hbar}} x\right).
\end{eqnarray}
Now, using Eqs.~(\ref{comp2}) and (\ref{ux}), coordinate space wave
function can be written as
\begin{eqnarray}\label{foo}
\psi(x)=\frac{1}{\sqrt{2\pi}}
\int_{-\frac{\pi}{2\sqrt{\beta}}}^{+\frac{\pi}{2\sqrt{\beta}}}
e^{\frac{i px}{\hbar}}\phi(p)\mathrm{d} p.
\end{eqnarray}
Moreover, $\phi(p)$ is given by the inverse Fourier transform of the
coordinate space wave function, namely
\begin{eqnarray}
\phi(p)=\frac{1}{\sqrt{2\pi}} \int_{-\infty}^{+\infty} e^{-\frac{i
px}{\hbar}}\psi(x)\mathrm{d} x.
\end{eqnarray}
To this end, by taking $\phi(p)=0$ for
$|p|>\frac{\pi}{2\sqrt{\beta}}$ we can formally extend the domain of
the momentum integral (\ref{foo}) to $-\infty<p<\infty$ without
changing the coordinate space wave function $\psi(x)$. Therefore,
$\psi(x)$ is the Fourier transform of $\phi(p)$. Now, using the
Babenko-Beckner inequality \cite{Babenko,Beckner} and following
Bialynicki-Birula and Mycielski \cite{bia1} we obtain (see
Ref.~\cite{bia2} for details)
\begin{eqnarray}\label{x+y}
S_x+S_p\geq1+\ln\pi,
\end{eqnarray}
where
\begin{equation}\label{Sxp}
S_x=-\int_{-\infty}^{+\infty}|\psi(x)|^2\ln |\psi(x)|^2\mathrm{d}x,
\hspace{2cm}S_p=-\int_{-\infty}^{+\infty}|\phi(p)|^2\ln |\phi(p)|^2\mathrm{d}p,
\end{equation}
subject to $\phi(p)=0$ for $|p|>\frac{\pi}{2\sqrt{\beta}}$. Note
that, in this representation, the expression for the entropic
uncertainty relation is similar to the ordinary quantum mechanics.
However, as we shall see in the next section, the presence of the
minimal length modifies the Hamiltonian, its solutions, and the
values of $S_x$ and $S_p$. But, since $\psi(x)$ and $\phi(p)$ are
still related by the Fourier transform, the lower bound for the
entropic uncertainty relation will not be modified.

\section{Quasi-position representation}
Another possible representation that exactly satisfies
Eq.~(\ref{gupc}) is \cite{4}
\begin{eqnarray}
X \phi(p) &=& i\hbar(1+\beta p^2)\partial_p\phi(p),\\ P\phi(p) &=&
p\phi(p).
\end{eqnarray}
The corresponding scalar product and completeness relations read
\begin{eqnarray}\label{comp1-2}
\langle p'|p\rangle=(1+\beta p^2) \delta(p-p'),\\
\int_{-\infty}^{+\infty}\frac{\mathrm{d} p}{1+\beta p^2}\,
|p\rangle\langle p|=1.\label{comp2-2}
\end{eqnarray}
Now, since the measure in the integral (\ref{comp2-2}) is not flat,
the momentum space entropy
\begin{equation}\label{ent1}
S_p=-\int_{-\infty}^{+\infty}\frac{\mathrm{d}p}{1+\beta
p^2}|\phi(p)|^2\ln |\phi(p)|^2,
\end{equation}
has no proper form of the continuous Shannon entropy relation in
this representation.

The quasi-position wave function is defined as \cite{4}
\begin{equation}
\phi({\xi}) \equiv\langle \psi^{ml}_{{\xi}} \vert \phi \rangle,
\end{equation}
where
\begin{equation}
\psi^{ml}_{\xi}(p) = \sqrt{\frac{2\sqrt{\beta}}{\pi}} (1+ \beta
p^2)^{-\frac{1}{2}} {\mbox{ }} e^{-i\frac{ {\xi}
\tan^{-1}(\sqrt{\beta}p)}{\hbar \sqrt{\beta}}},
\end{equation}
denotes the maximal localization states. These states satisfy
$\langle \psi^{ml}_{{\xi}}\vert X \vert \psi^{ml}_{{\xi}}\rangle =
{\xi} $, $(\Delta X)_{\vert \psi^{ml}_{{\xi}} \rangle } =
\hbar\sqrt{\beta}$, and are not mutually orthogonal, i.e., $\langle
\psi^{ml}_{{\xi}^{\prime}} \vert \psi^{ml}_{{\xi}} \rangle
\ne\delta(\xi-\xi')$. In this representation, the relation between
the momentum space wave functions and  quasi-position wave functions
is given by
\begin{equation}
\psi({\xi}) = \sqrt{\frac{2\sqrt{\beta}}{\pi}}
\int_{-\infty}^{+\infty} \frac{dp}{(1+\beta p^2)^{3/2}}
e^{\frac{i{\xi} \tan^{-1}(\sqrt{\beta} p)}{ \hbar \sqrt{\beta}}}
\psi(p).
\end{equation}
Thus, the quasi-position wave functions are not Fourier transform of
the momentum space wave functions and the quasi-position entropy
\begin{equation}\label{ent2}
S_\xi=-(8\pi\hbar^2\sqrt{\beta})^{-1} \int_{-\infty}^{+\infty}
\int_{-\infty}^{+\infty} \int_{-\infty}^{+\infty} dp {\mbox{ }}
d{\xi} {\mbox{ }} d{\xi}^\prime {\mbox{ }}
e^{i({\xi}-{\xi}^\prime)\frac{\tan^{-1}(\sqrt{\beta}p)}{\hbar
\sqrt{\beta}}}{\mbox{ }} \psi^*({\xi})\psi({\xi}^\prime)\ln
|\psi(\xi')|^2 ,
\end{equation}
does not represent the continuous Shannon entropy and contains
plenty of overcounting (because of the non-orthogonality of $\vert
\psi^{ml}_{{\xi}} \rangle$) that should be avoided.
These results show that the information entropies $S_p$ (\ref{ent1})
and $S_\xi$ (\ref{ent2}) are not proper measures of uncertainty.
However, the information entropies (\ref{Sxp}) based on formally
self-adjoint representation do not suffer from these shortcomings
and they can be considered as proper measures of uncertainty in the
presence of the minimal length.

\section{Quantum oscillator}
The Hamiltonian of the harmonic oscillator is given by
$H=\frac{P^2}{2m}+\frac{1}{2}m\omega^2X^2$.  So, the generalized
Schr\"odinger equation in momentum space using the representation
(\ref{x0p01},\ref{x0p02}) reads
\begin{eqnarray}
-\frac{1}{2}m\hbar^2\omega^2\frac{d^2\phi(p)}{dp^2}+\frac{\tan^2\left(\sqrt{\beta}p\right)}{2m\beta}\phi(p)=E\phi(p).
\end{eqnarray}
Using the new variable $\xi=\sqrt{\beta}p$, the above equation can
be written as
\begin{eqnarray}
\frac{d^2\phi(\xi)}{d\xi^2}+\left(\epsilon-\frac{V}{\cos^2\xi}\right)\phi(\xi)=0,
\end{eqnarray}
where $V=\left(m\beta\hbar\omega\right)^{-2}$ and
$\epsilon=V(1+2m\beta E)$. Now, taking
\begin{eqnarray}
\phi_n(\xi)=P_n(s)\cos^\lambda \xi,
\end{eqnarray}
results in
\begin{eqnarray}
(1-s^2)\frac{d^2P_n(s)}{ds^2}-s(1+2\lambda)\frac{dP_n(s)}{ds}+(\epsilon-\lambda^2)P_n(s)=0,
\end{eqnarray}
where $s=\sin\xi$ and
\begin{eqnarray}
V=\lambda(\lambda-1),\hspace{2cm}\lambda=\frac{1}{2}\left[1+\sqrt{1+\frac{4}{m^2\beta^2\hbar^2\omega^2}}\,\right].
\end{eqnarray}
It is known that the solutions of the above equation for
$\epsilon=(n+\lambda)^2$ are given by the Gegenbauer polynomials
$C^{\lambda}_n(s)$. Therefore, the exact solutions read
\begin{eqnarray}\label{solg}
\phi_n(p)&=&N_n\,C^{\lambda}_n\left(\sin(\sqrt{\beta}p)\right)\cos^\lambda (\sqrt{\beta}p),\\
E_n&=&\hbar\omega\left(n+\frac{1}{2}\right)\left(\sqrt{1+\eta^2/4}+\eta/2\right)+\frac{1}{2}\hbar\omega\eta
n^2,\hspace{1cm}n=0,1,2,...,
\end{eqnarray}
where $\eta=m\beta\hbar\omega$, $N_n$ is the normalization
coefficient, and the Gegenbauer polynomials are defined as \cite{32}
\begin{eqnarray}
C^{\lambda}_n(s)=\sum_{k=0}^{[n/2]}(-1)^k\frac{\Gamma(n-k+\lambda)}{\Gamma(\lambda)k!(n-2k)!}(2s)^{n-2k}.
\end{eqnarray}
Note that for $\beta\rightarrow0$ we obtain the ordinary energy
spectrum of the harmonic oscillator, i.e.,
$E_n=\hbar\omega\left(n+\frac{1}{2}\right)$. The Gegenbauer
polynomials also satisfy the following useful formula \cite{32}
\begin{eqnarray}
\int_{-1}^1
(1-x^2)^{\nu-\frac{1}{2}}[C^{\nu}_n(x)]^2dx=\frac{\pi2^{1-2\nu}\Gamma(n+2\nu)}{n!(n+\nu)[\Gamma(\nu)]^2},\hspace{1cm}\mathrm{Re}\,\nu>-\frac{1}{2}.
\end{eqnarray}
Since $s=\sin\xi$ and $N_n$ is given by the normalization condition
$\int_{-\frac{\pi}{2\sqrt{\beta}}}^{+\frac{\pi}{2\sqrt{\beta}}}\mathrm{d}p\,|\phi(\xi)|^2=1$,
we find
\begin{eqnarray}
N_n=\sqrt{\frac{\sqrt{\beta}\,n!(n+\lambda)[\Gamma(\lambda)]^2}{\pi2^{1-2\lambda}\Gamma(n+2\lambda)}}.
\end{eqnarray}
The solutions can be also written in terms of relativistic Hermite
polynomials using the relation \cite{Nagel}
\begin{eqnarray}
H_n^\lambda(\sqrt{\lambda}u)=\frac{n!}{\lambda^{n/2}}(1+u^2)^{n/2}C_n^\lambda\left(\frac{u}{\sqrt{1+u^2}}\right),
\end{eqnarray}
where $H_n^\lambda(z)$ denotes relativistic Hermite polynomials.
Thus, we obtain
$C_n^\lambda\left(\sin(\sqrt{\beta}p)\right)=\frac{\lambda^{n/2}}{n!}\cos^n(\sqrt{\beta}p)H_n^\lambda\left(\sqrt{\lambda}\tan(\sqrt{\beta}p)\right)$
which results in
\begin{eqnarray}
\phi_n(p)=\frac{N_n\lambda^{n/2}}{n!}\cos^{\lambda+n}(\sqrt{\beta}p)H_n^\lambda\left(\sqrt{\lambda}\tan(\sqrt{\beta}p)\right).
\end{eqnarray}
For the small values of the deformation parameter we have
($m=\hbar=1$)
\begin{eqnarray}
&&\lim_{\beta\rightarrow0}\lambda=\frac{1}{\omega\beta},\hspace{2cm}
\beta\rightarrow0\Longleftrightarrow\lambda\rightarrow\infty,\\
&&\lim_{\lambda\rightarrow\infty}\cos^{\lambda+n}(\sqrt{\beta}p)=\exp\left(-\frac{p^2}{2\omega}\right),\\
&&\lim_{\lambda\rightarrow\infty}\sqrt{\lambda}\tan(\sqrt{\beta}p)=\frac{p}{\sqrt{\omega}},\\
&&\lim_{\lambda\rightarrow\infty}H_n^\lambda(z)=H_n(z),\hspace{2cm}\mbox{Ref.~\cite{Nagel}},
\end{eqnarray}
where $H_n$ denotes Hermite polynomials. So, in this limit the
solutions read
\begin{eqnarray}
\lim_{\beta\rightarrow0}\phi_n(p)=\frac{e^{-\frac{p^2}{2\omega}}H_n\left(\frac{p}{\sqrt{\omega}}\right)}{\sqrt{2^nn!\sqrt{\pi\omega}}},
\end{eqnarray}
which are normalized eigenstates of the ordinary harmonic oscillator
as we have expected.

\section{Information entropy}
The information entropies for the position and momentum spaces can
be now calculated for the harmonic oscillator in the GUP framework
using Eq.~(\ref{Sxp}). In ordinary quantum mechanics and in the
position space, the information entropy can obtained analytically
for some quantum mechanical systems. However, since the momentum
wave functions are derived from the Fourier transform, the
corresponding momentum information entropies are rather difficult to
obtain. For our case, as we shall show, we find $S_p$ analytically
for the two lowest energy states and obtain $S_x$ numerically. Thus,
because of the difficultly in calculating $S_x$ and $S_p$, we only
consider the two lowest energy eigenstates.

\begin{figure}
\centerline{\begin{tabular}{ccc}
\includegraphics[width=8cm]{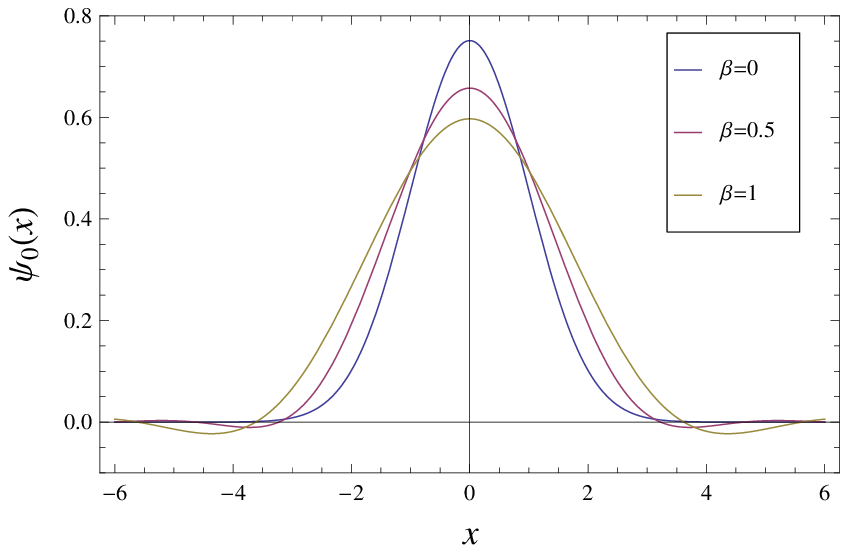}
 &\hspace{2.cm}&
\includegraphics[width=8cm]{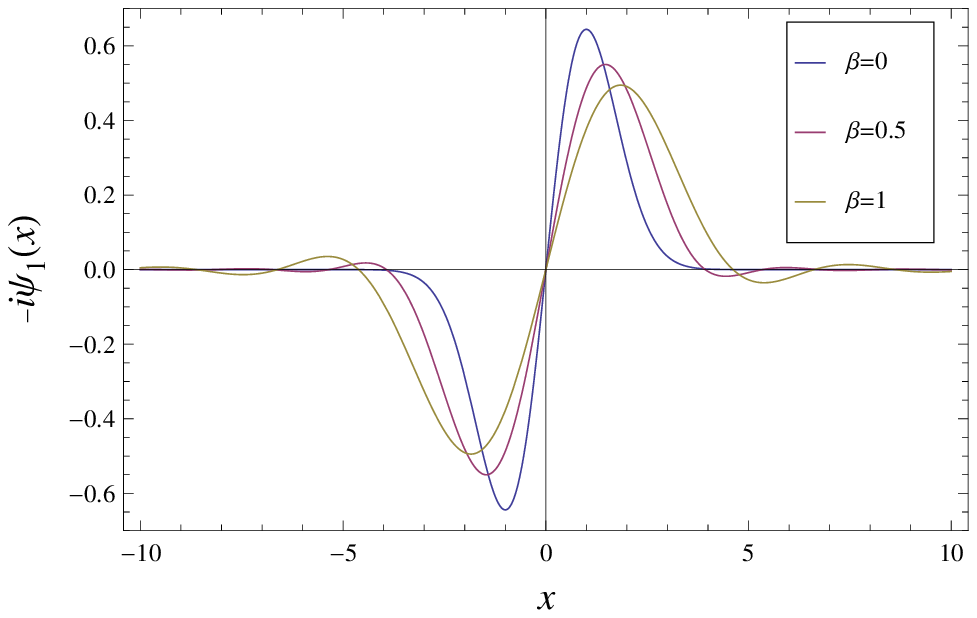}
\end{tabular}}
\caption{\label{fig1} Plots of the position space wave functions for
$m=\hbar=\omega=1$ and $n=0,1$.}
\end{figure}

First consider the Fourier transform of the momentum space ground
state (\ref{solg}) which gives the following state in the position
space ($\hbar=1$)
\begin{eqnarray}
\psi_0(x)=\sqrt{\frac{\lambda[\Gamma(\lambda)]^2}{\pi^2\sqrt{\beta}\Gamma(2\lambda)}}\frac{\sin\left[\frac{\pi}{2}
\left(\frac{x}{\sqrt{\beta}}-\lambda\right)\right]}{\frac{x}{\sqrt{\beta}}-\lambda}\,
_2F_1 \left[\frac{1}{2}
\left(\frac{x}{\sqrt{\beta}}-\lambda\right),-\lambda,\frac{1}{2}
\left(\frac{x}{\sqrt{\beta}}-(\lambda-2)\right),1\right].
\end{eqnarray}
Also, for the first excited state ($n=1$) we have
\begin{eqnarray}
\psi_1(x)&=&i\lambda\sqrt{\frac{(\lambda+1)[\Gamma(\lambda)]^2}{\pi^2\sqrt{\beta}\Gamma(2\lambda+1)}}
\Bigg\{\frac{\sin\left[\frac{\pi}{2}
\left(\frac{x}{\sqrt{\beta}}-(\lambda+1)\right)\right]}{\frac{x}{\sqrt{\beta}}-(\lambda+1)}\nonumber\\
&&\times_2F_1 \left[\frac{1}{2}
\left(\frac{x}{\sqrt{\beta}}-(\lambda+1)\right),-\lambda,\frac{1}{2}
\left(\frac{x}{\sqrt{\beta}}-(\lambda-1)\right),1\right]-\frac{\sin\left[\frac{\pi}{2}
\left(\frac{x}{\sqrt{\beta}}-(\lambda-1)\right)\right]}{\frac{x}{\sqrt{\beta}}-(\lambda-1)}\nonumber\\
&&\times _2F_1 \left[\frac{1}{2}
\left(\frac{x}{\sqrt{\beta}}-(\lambda-1)\right),-\lambda,\frac{1}{2}
\left(\frac{x}{\sqrt{\beta}}-(\lambda-3)\right),1\right]\Bigg\}.
\end{eqnarray}
Figure \ref{fig1} shows the resulting ground state and first excited
state in position space for $\beta=\{0,0.5,1\}$. Notice that, for
$\beta\rightarrow0$ the solutions tend to the simple harmonic
oscillator wave functions, i.e.,
$\lim_{\beta\rightarrow0}\psi_n(x)=\sqrt{\frac{\sqrt{\omega}}{2^nn!\sqrt{\pi}}}e^{-\omega
x^2/2}H_n\left(\sqrt{\omega}x\right)$.

For the ground state, the analytical expression for $S_p$ reads
\begin{eqnarray}
S^0_p=\lambda H_{\lambda}-\lambda
H_{\lambda-\frac{1}{2}}+\ln\sqrt{\pi}-\ln\left(\frac{\sqrt{\beta}\lambda\Gamma(\lambda)}{\Gamma(\lambda+\frac{1}{2})}\right),
\end{eqnarray}
where $H_\lambda$ denotes the harmonic number $H_n=\sum_{k=1}^n
1/k$. It is easy to check that for the small values of $\beta$, the
momentum information entropy tends to the ordinary harmonic
oscillator information entropy, namely
\begin{eqnarray}
\lim_{\beta\rightarrow0}S^0_p=\lim_{\beta\rightarrow0}S^0_x=\frac{1}{2}\left(1+\ln
\pi\right).
\end{eqnarray}
For the first excited state, $S_p$ is given by
\begin{eqnarray}
S^1_p=(1+\lambda) H_{\lambda+1}-\lambda
H_{\lambda-\frac{1}{2}}+\ln\sqrt{\pi}-2-\ln\left(\frac{\sqrt{\beta}\Gamma(\lambda+2)}{2\Gamma(\lambda+\frac{1}{2})}\right),
\end{eqnarray}
and for $\beta\rightarrow0$ it reads
\begin{eqnarray}
\lim_{\beta\rightarrow0}S^1_p=\lim_{\beta\rightarrow0}S^1_x=\frac{1}{2}\ln
\pi+\ln2+\gamma-\frac{1}{2},
\end{eqnarray}
where $\gamma\approx0.5772$ is the Euler constant.

\begin{figure}
\centerline{\begin{tabular}{ccc}
\includegraphics[width=8cm]{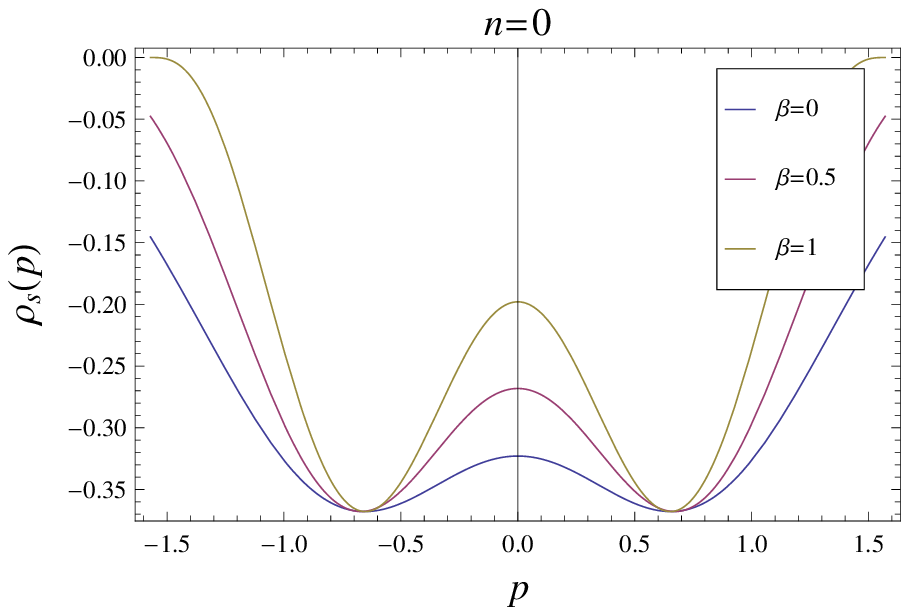}
 &\hspace{2.cm}&
\includegraphics[width=8cm]{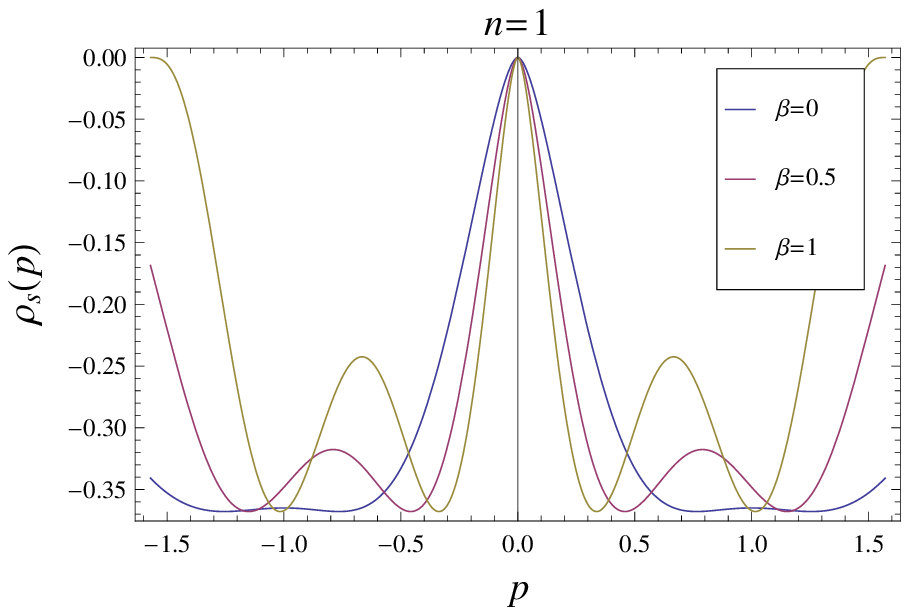}
\end{tabular}}
\caption{\label{fig2} Plots of the momentum space entropy densities
for $m=\hbar=\omega=1$ and $n=0,1$.}
\end{figure}

\begin{figure}
\centerline{\begin{tabular}{ccc}
\includegraphics[width=8cm]{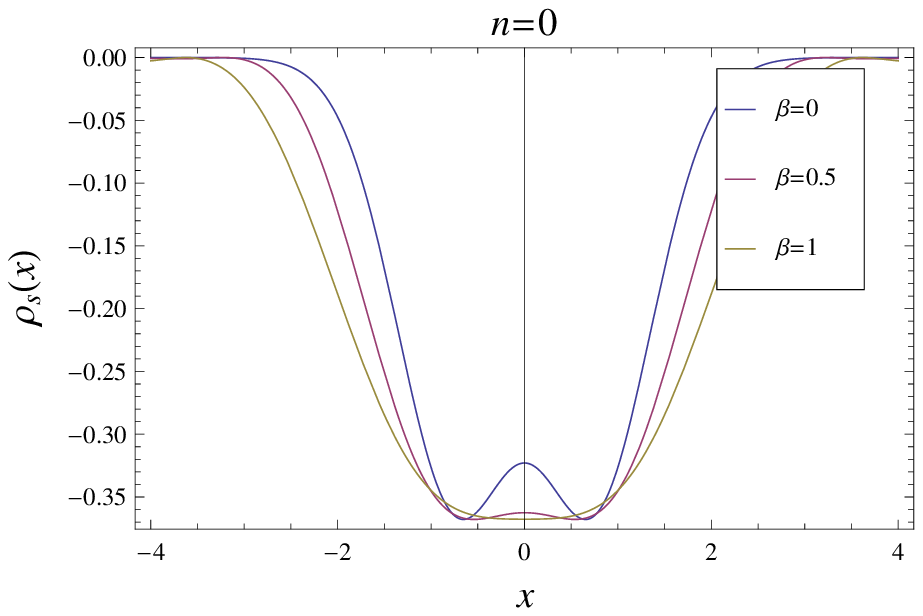}
 &\hspace{2.cm}&
\includegraphics[width=8cm]{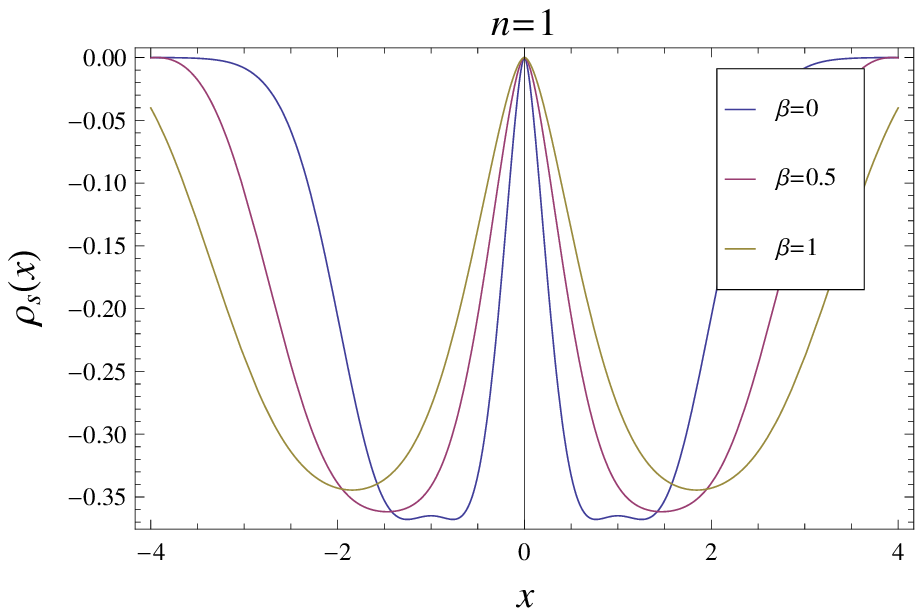}
\end{tabular}}
\caption{\label{fig3} Plots of the position space entropy densities
for $m=\hbar=\omega=1$ and $n=0,1$.}
\end{figure}

\begin{table}
\caption{\label{tab1}The numerical results that establish the BBM
entropic uncertainty relation for $m=\hbar=\omega=1$ and various
values of $\beta$.}
\begin{ruledtabular}
\begin{tabular}{llllll}
$n$ & $\beta$ & $S_x$ & $S_p$ & $S_x+S_p$& $1+\ln\pi$\\ \hline
0 & 0.1 & 1.12153  & 1.02361 & 2.14515  & 2.14473\\
 & 0.5 & 1.30251  & 0.85220 & 2.15471  &  2.14473\\
 & 1.0 &  1.49095 & 0.68153 & 2.17248  &    2.14473\\
 1 & 0.1 & 1.40656  &1.24992 & 2.65648  & 2.14473\\
 & 0.5 & 1.62672  & 0.97566 & 2.60238  &  2.14473\\
 & 1.0 &  1.84423 &0.74892 &  2.59315  &    2.14473\\
\end{tabular}
\end{ruledtabular}
\end{table}

The information entropy densities are defined as
$\rho_s(x)=|\psi(x)|^2\ln|\psi(x)|^2$ and
$\rho_s(p)=|\phi(p)|^2\ln|\phi(p)|^2$. The behavior of $\rho_s(x)$
and $\rho_s(p)$ is illustrated in Figs.~\ref{fig2} and \ref{fig3}
for $n=0,1$ and several values of the deformation parameter. Now,
using the numerical values for $S_x$, we obtain the left hand side
of Eq.~(\ref{x+y}). In Table \ref{tab1}, we have reported the
position and momentum space information entropies for
$\beta=\{0.1,0.5,1\}$ and showed that they obey the BBM inequality.
These results indicate that the position space information entropy
increases with the GUP parameter $\beta$ and vice versa for the
momentum space information entropy but their sum stays above the
value $1+\ln\pi$.

\section{Conclusions}
In this paper, we studied the Shannon entropic uncertainty relation
in the presence of a minimal measurable length proportional to the
Planck length. We showed that using the formally self-adjoint
representation, since the coordinate space and momentum space wave
functions are related by the Fourier transformation, the measure in
the information entropic integral is flat and the lower bound that
is predicted by the BBM inequality is guaranteed in the GUP
framework. It is worth mentioning that the BBM inequality does not
hold for all wave functions. In fact, its validity depends on both
the deformed algebra and its representation. As we have indicated,
this inequality does not hold in quasi-position representation of
our algebra $[X,P]=i\hbar\left(1+\beta P^2\right)$. As another
example, we mentioned the algebra $[X,P]=i\hbar\left(1+\alpha
X^2+\beta P^2\right)$ that implies both a minimal length and a
minimal momentum. Since this algebra has no formally self-adjoint
representation in the form $X=x$ and $P=f(p)$, the momentum space
and coordinate space wave functions are not related by the Fourier
transform and the BBM uncertainty relation is not valid for this
form of GUP. For the case of the harmonic oscillator, we exactly
solved the generalized Schr\"odinger equation in momentum space and
found the solutions in terms of the Gegenbauer polynomials. Also,
for the two lowest energy eigenstates, we obtained the solutions in
position space in terms of the hypergeometric functions. Then, the
analytical expressions for the information entropies are found in
the momentum space with proper limiting values for
$\beta\rightarrow0$. Using the numerical values for the position
information entropy, we explicitly showed that the BBM inequality
holds for various values of the deformation parameter. To check the
validity of the BBM inequality for other potentials, we need to
solve the generalized Schr\"odinger equation which contains higher
order differential terms. However, for the small anharmonic
potential terms, the perturbation theory can be used to find the
approximate solutions. Also, For other types of GUPs, if a formally
self-adjoint representation is viable, the BBM inequality is still
valid.

\end{document}